\documentstyle[multicol,prl,aps,epsf,psfig,epsfig]{revtex}

\begin{document}

\title
{
Self-organized random walks and stochastic sandpile: \\
From linear to branched avalanches
}

\author
{
S. S. Manna$^{(1,2)}$ and A. L. Stella$^{(1,3)}$
}

\address
{
$^{(1)}$ INFM - Dipartimento di Fisica, Universit\`a di Padova, I-35131
Padova, Italy \\
$^{(2)}$Satyendra Nath Bose National Centre for Basic Sciences,
    Block-JD, Sector-III, Salt Lake, Kolkata-700098, India \\
$^{(3)}$ Sezione INFN, Universit\`a di Padova, I-35131 Padova, Italy
}
\maketitle
\begin{abstract}

In a model of self-organized criticality unstable sites discharge 
to just one of their neighbors. For constant discharge ratio $\alpha$ and
for a certain range of values of the input energy,
avalanches are simple branchless P\'olya random walks, and their
scaling properties can be derived exactly. If $\alpha$ fluctuates widely enough,
avalanches become branched, due to multiple discharges, and behave like
those of the stochastic sandpile. At the threshold for branched behaviour,
peculiar scaling and anomalous diffusive transport are observed.

PACS 05.40.+j, 05.70.Jk, 05.70.Ln
\end{abstract}

\begin{multicols}{2}

   An avalanche is a cascade of a large number of microscopic 
events. Generally it is triggered by a single event at a point.
This creates similar events in the neighbourhood
which activate the neighbours at further distances. Thus, an
avalanche is a simple example of branched growth process,
of which several physical examples exist\cite{avalanches}. 
Particularly interesting are situations in which
the avalanche sizes have a scale free distribution, typically 
a simple power law, signifying
the existence of long-range correlations as in critical
phenomena. This is the case of
avalanches in the phenomenon of Self-Organized Criticality 
(SOC) \cite {btw} where long-ranged spatio-temporal correlations 
spontaneously emerge in non-equilibrium
steady states of slowly driven systems with nonlinear
local relaxation mechanisms\cite {socbook,soc}. 

   Sandpiles are prototype models of SOC. $n_i$ grains reside 
at the $i$-th site of a regular lattice. The role of the external drive
is to trigger transport processes through the system by adding single
grains of sand at a time at randomly selected sites: 
$n_i \to n_i+1$. If $n_i \ge n^c$ the sand column at
$i$ topples and grains are distributed to the neighbourhood. 
In the BTW model, 
the grain distribution process is deterministic since each neighbouring
site gets one grain \cite {btw}. 
In the stochastic two-state sandpile model 
grains are transferred to randomly chosen neighbouring
sites \cite {manna}. 
In spite of their similarities, the BTW model and the stochastic
sandpile are now believed to belong to different universality
classes\cite{biham}. Indeed, very recent studies show that
the BTW model has a multifractal behaviour\cite{stella1,stella2}, 
whereas standard finite size scaling works well for the two-state
stochastic sandpile\cite{stella2,lubeck,chessa,vesp}. 

    Characterizing features of all these sandpile models are (i) the normal diffusive 
dynamics of the particles and (ii) the branching of the toppling process. A grain moves
a unit distance in a toppling and its resultant motion under different
topplings in different avalanches is diffusive. This implies that the
average number of topplings in an avalanche grows as a quadratic power of the
system size, $L$, in all isotropic sandpile models. Secondly in all models, the toppling
condition is made in such a way that a single toppling can excite more
than one neighbors, which ensures that an avalanche is a branched process. 
Normal diffusive transport and
branching are believed to be very basic ingredients of SOC.

   In this paper, by studying an original energy activation model with stochastic
discharge mechanism, we find that neither of these
ingredients (i) and (ii) as stated above, is necessary for SOC behaviour. 
Indeed, we show that avalanches in SOC 
can be linear as branchless random walks. In this case their scaling behaviour
is different from that of branched avalanches, 
Eulerian random walk models are also linear avalanche models studied before \cite {ERW}.
and the relative probability distributions do not become
independent of $L$, as this tends to
infinity. By allowing the discharge ratio to fluctuate in a
progressively wider interval, we are also
able to trigger a branched avalanche behaviour, which falls in the universality
class of the stochastic sandpile\cite{manna}. Right at threshold
for this branched behaviour the diffusive transport
becomes anomalous, i.e. the average avalanche size grows 
as a power higher than $2$ of $L$.

   Our model has two parameters: the input amount $\delta$ and 
the discharge ratio $\alpha$.
An amount of energy $\epsilon_i$, which can vary continuously,
is accumulated at each site $i$ of a
square lattice box of size $L$.
The system is driven by injecting an amount of energy
$\delta$ to a randomly selected site. Every lattice site
has a limiting capacity $\epsilon_c(=1)$ for the maximum amount of energy storage.
If at any site the energy $\epsilon_i > \epsilon_c$, the site $i$ activates and undergoes a 
relaxation process.
In a relaxation a fraction $\alpha$ of the site energy is
discharged to only {\it one} of the neighbouring sites, $j$,
which is selected randomly: $\epsilon_j \rightarrow \epsilon_j + \alpha \epsilon_i$. 
The rest of the energy remains in the relaxing site: 
$\epsilon_i \rightarrow (1-\alpha)\epsilon_i$. If the energy at the
receiving site $j$ exceeds the threshold, that site also relaxes.
The possibility of multiple relaxations arises when site $i$ remains
active even after relaxing. In such a case further relaxations occur
in the subsequent stages. Upon relaxing, an active site at the 
boundary may drop a fraction $\alpha$ of its 
energy outside the system. This ensures that the system reaches
unique 
\end{multicols}
\begin{figure}[top]
\begin{center}
\includegraphics[width=12cm]{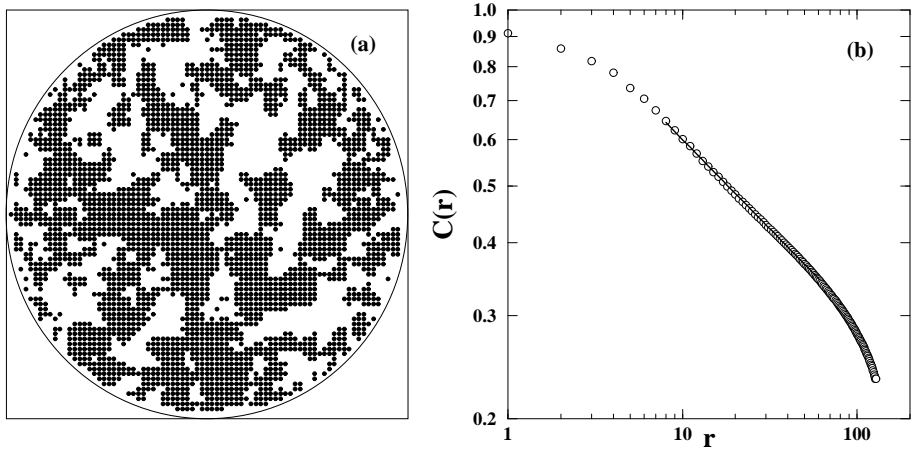}
\end{center}
\caption{
(a) A steady state energy configuration in the LAM ($\alpha=1/3$, $\delta=1/5$) within
a circular region on a square lattice of size $L=81$.
(b) Plot of the correlation function $C(r)$ for the same LAM as in (a) but for $L$ = 257.
}
\label{fig1}
\end{figure}
\begin{multicols}{2}
stationarity after repeated driving. Our simulations show that 
this is the case, independent of starting configuration as required in SOC,
for any particular selection of ($\alpha$,$\delta$). We note that this model 
is a special case of the class of models called the abelian distributed 
processors model \cite {Dhar1}.

   First we consider a constant value of $\alpha$.
In the steady state, $1-\alpha \le \epsilon_i \le 1$ since each site 
relaxes at least once. This implies that a site must activate after receiving at least 
energy $\alpha$ due to an activation at the neighbouring site.
Energy activation events then take place one after 
another like in a step sequence of a P\'olya random walk (RW) \cite {Polya}. 
Once started, such a RW must terminate at the boundary, after
dropping some energy outside the system. We call this the
linear avalanche model (LAM).

   When $\delta \le \alpha$ there must be some sites which do not activate upon 
receiving the input energy, since a RW drops at least $\alpha$ energy 
outside the system. To eliminate corner effects we 
use a circular region, within a square box of size $L$ (odd), of radius $R=(L-1)/2$.
Fig. 1(a) shows an energy configuration in the steady state where the sites
having energy larger than the average, $\langle \epsilon \rangle$, are represented by
black dots.
Correlated regions of such sites are observed as connected clusters of black dots.
The correlation function $C(r)$ is defined as the probability that
a black dot at a distance $r$ from the centre belongs to the same
cluster containing a black dot at the centre. We observe a power law decay
$C(r) \sim r^{-\xi}$ with $\xi = 0.37 \pm 0.02$ which is the signature of the
critical correlation developed in the steady state (Fig. 1(b)).
However a site energy correlation function like: 
$\langle e(0)e(r)\rangle-\langle e^2 \rangle$ averaged
over all sites has an almost uniform small negative value except when $r \sim R$.

   The probability density of the site energies $D(\epsilon,R)$
in the steady state is very well
fitted with a generalized Lorentzian peak in the range $1- \alpha < \epsilon < 1$
around its average $\langle \epsilon \langle $. $D$ follows a scaling with $R$ of the form:
\begin{equation}
D(\epsilon)/R=a^2/\{(\epsilon- \langle \epsilon \rangle)^2R^2+b^2\}^c
\end{equation}
where, for example, $a \approx 0.195, b \approx 0.226$ and $c \approx 1.33$, with $\alpha=1/3$
and $\delta=1/5$ (Fig. 2). From Fig. 2(a) we see that the site energy distribution
is symmetric about its average value at $\langle \epsilon \rangle$ and the most probable value
coincides with the average value. Assuming that the $\epsilon>$ is less than
$1-\delta$, the fraction of sites which donot activate on receiving the input energies
are greater than those which activate. This situation cannot be stable since
these low energy $(\epsilon < 1-\delta)$ sites absorb energy at a greater
rate from the external drive than the high energy sites $(\epsilon > 1-\delta)$
and the average energy will push up. Similarly if  $\langle \epsilon \rangle$ is greater
than $1-\delta$ there would be faster dissipation of energy through the
boundary which will push down the average energy. Therefore it is likely
that the stable states correspond to $\langle \epsilon \rangle = 1-\delta$ where the rates of
absorption by the low energy sites and the rate of dissipation by the
inputs at the high energy sites are equal. Our numerical results strongly
supports this result:
\begin{equation}
\langle \epsilon(\alpha,\delta) \rangle = 1- \delta \quad {\rm for} \quad \delta \le \alpha
\end{equation}

   For $ \alpha \le \delta \le \delta_c(\alpha)$
every RW drops exactly an amount of energy
$\delta$ outside the boundary. $D(\epsilon,R)$ is a delta function
at its average value $\langle \epsilon \rangle$ such that a site after activation 
retains the same energy which it had before receiving the external input
$\delta$. This implies
\begin{equation}
(1-\alpha)(\langle \epsilon \rangle+\delta)=\langle \epsilon \rangle
\end{equation}
which gives $\langle \epsilon(\alpha,\delta) \rangle = (1/\alpha-1)\delta$.
The average energy increases to 1 when the input energy is increased to 
$\delta_c(\alpha)=\alpha/(1-\alpha)$. This is the limiting situation
for the branchless avalanches. Beyond this limit multiple discharges
start and avalanches cease to be branchless. 

\end{multicols}
\begin{figure}[t]
\begin{center}
\includegraphics[width=12cm]{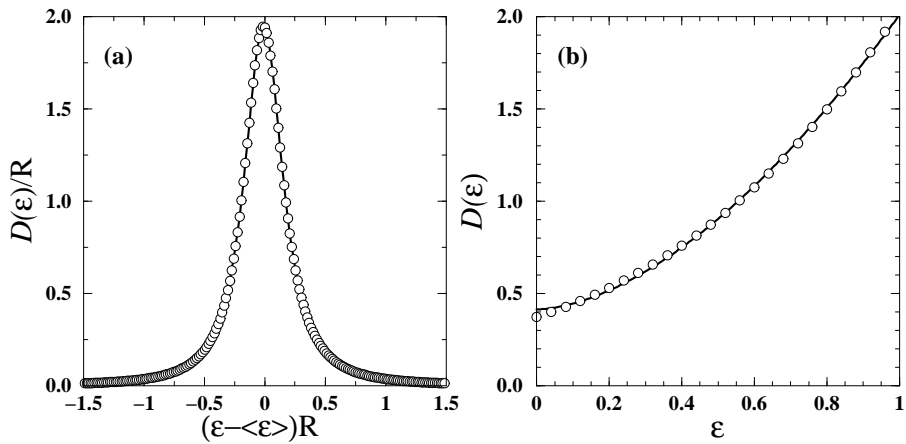}
\end{center}
\caption{
$D(\epsilon)$ for
(a) LAM and (b) BAM. The continuous curves are fits to a generalized Lorentzian function in (a) and
to a power law in (b).
}
\label{fig2}
\end{figure}
\begin{multicols}{2}

   The size of the avalanche is generally measured by the
total number of relaxations $(s)$, the life-time of the avalanche ($t$) and the radius 
of the avalanche $(r)$.  Using $\{x: s, t, r\}$ we assume 
$x \sim x'^{\gamma_{x'x}}$, where the $\gamma_{rx}$ is the cut-off exponent for 
the measure $x$, 
and the usual finite size scaling forms
for the probability distribution functions of avalanche
sizes:
$
{\rm Prob}(x,L) \sim L^{-\beta_x} f_x \left (\frac {x}{L^{\gamma_{rx}}}\right ), \quad
$
where, $f_x(y) \sim y^{-\tau_x}$ in the limit of $y \to 0$
gives $\tau_x = \beta_x/\gamma_{rx}$,
in case ${\rm Prob}$ does not maintain an $L$ dependence
for $s<<L$. The $\gamma$ exponents are connected by the relations:
$\gamma_{x'x} = \gamma_{x'x''} \gamma_{x''x}$
and $\gamma_{xx'}=(\tau_{x'}-1)/(\tau_x-1)$.

   The steady state of the LAM is related to the first passage problem of RW's.
The avalanche size and the life-time are the same as the number of steps $s$ taken
by the walker before dropping through the boundary. The distribution of the avalanche   
sizes can be calculated in the following way. The probability to start a RW within a 
distance $\Delta R$ from the boundary is $2\pi R \Delta R / \pi R^2 \sim \Delta R / R$.
A RW moves a distance $\sim s^{1/2}$ in $s$ steps. Therefore, the probability that
an arbitrary walker makes $s$ steps or less before reaching the boundary is the probability
that an arbitrary site is within a distance $s^{1/2}$ from the boundary, which is
$ \sim s^{1/2} / R$. The probability that an arbitrary walker makes 
precisely $s$ steps 
to reach the boundary is $s^{-1/2} / R$. Thus, in terms of $L$,
Prob$(s,L) \sim s^{-\tau_s} / L$ for $s << L$ with $\tau_s=1/2$. Our 
simulations show that upon increasing the system size the effective exponent $\tau_s(L)$ 
gradually decreases to its asymptotic limit of $0.50(1)$.
Since the avalanches are random walks, the cut-off for $s$ must
be proportional to $L^2$, and therefore $\gamma_{rs}=2$. Normalization of Prob$(s,L)$ gives
$\beta_s=2$. 
The relation $\tau_s=\beta_s/\gamma_{rs}$ is not satisfied by the LAM avalanches,
because ${\rm Prob}$ maintains an $L$ dependence for $s<<L$. This unusual dependence is due
to the fact that LAM avalanches must necessarily reach the boundary in order to
extinguish. As a rule, this is not the case for other SOC systems.

   We compare these results with two cases studied in the literature.
A model of SOC where avalanches are branchless walks is the 
Eulerian walkers model. In this model each site of a regular lattice
has an outgoing direction, which is one member of a set of outgoing bonds
associated with this site. The walker leaves the site along the outgoing direction
but changes the outgoing direction sequentially to the next member
in the set of outgoing bonds \cite {ERW}. 
In the Euler walker case, the dependence of Prob$(s,L)$ is of the form 
$L^{-2} f(sL^{-2})$ just similar to our LAM where $s$ is the number of steps
to the boundary. Secondly for the random walks
on a square with absorbing boundary, the exact scaling function for the distribution of
number of steps to the boundary is $P(s) \sim L^{-2}f(sL^{-2})$ has been 
calculated. It has been shown that the scaling function $f(x)$ can be
expressed explicitly in terms of the Jacobi theta function \cite {Pradhan}.

   Next we study a situation when the discharge ratio $\alpha$ is a random variable and 
a fresh value for it is drawn from a uniform random distribution in $\{0,1\}$ each time a 
site relaxes. Since now the energy released after activation may be arbitrarily small,
multiple relaxations are quite frequent, and this leads to branching. 
We call this the branched avalanche model (BAM). 
The probability density $D(\epsilon)$ of the site energies in the steady state has     
now little dependence on $\delta$ as well as on $L$.
$D(\epsilon)$ is fitted best to a form $D(\epsilon) = D_o + D_1\epsilon^{\mu}$, with
$D_o=0.41(1)$, $D_1=1.59(1)$ and $\mu=1.70(2)$ (Fig. 2).
The average energy per site $\langle \epsilon \rangle$ is however observed to have an
$L$ dependence as: $\langle \epsilon \rangle = \epsilon_{\infty} - CL^{-3/4}$ with $\epsilon_{\infty} = 0.6365(3)$ and
$C=0.21(2)$.

   The probability distributions for avalanche sizes and life-times
follow the finite size scaling form of Prob very well.
The data for Prob$(s,L)$ for three different system sizes $L$ = 129, 513 and 2049
collapse well for $\gamma_{rs}=2.75$ and
$\beta_s=3.6$. This gives a value for $\tau_s=1.31$. 
Similarly, we obtain $\tau_t$ = 1.51(3) and
$\gamma_{rt}$ = 1.50(3). The average size $\langle s(L) \rangle \sim L^{\nu_s}$
with $\nu_s=2$ and similarly $\nu_t=0.76(3)$ are obtained.
The values of these exponents are very close to those of
the two-state model, indicating that the BAM may well belong to the
universality class of the two-state model \cite{manna}.
In the BAM, avalanches can extinguish within
the boundary.

\begin{figure}[t]
\begin{center}
\includegraphics[width=6cm]{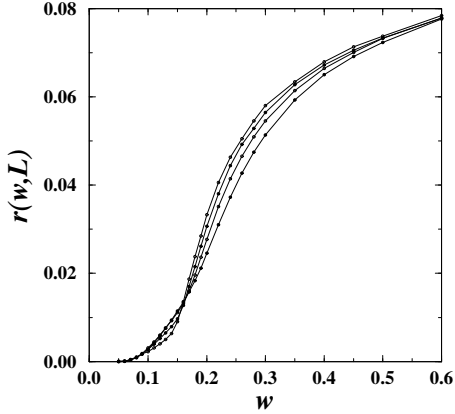}
\end{center}
\caption{
Variation of the average branching number $r(w,L)$ with $w$. The curves
become steeper as the system size $L$ increases from 33 to 512. The
threshold for BAM is at $w_c = 0.125 \pm 0.010$.
}
\label{fig3}
\end{figure}

The LAM and the BAM can be regarded as the two extremes
of a range of situations characterized by a progressively increasing 
width of the interval in which $\alpha$ is allowed to fluctuate. 
We introduce a parameter
$w$ as the size of the window for the stochastic $\alpha$, 
which is now chosen randomly within the range $\{1/2-w/2,1/2+w/2\}$ 
with uniform probability. When $w=0$ we have the LAM with a constant 
$\alpha=1/2$, whereas when $w=1$ we have the BAM. 
A full characterization of all dynamical regimes for $0<w<1$ turns out
to be extremely challenging. However, starting from $w=1$ and 
progressively lowering this parameter, we are able to identify a whole region
$w_c<w<1$ in which BAM behaviour holds, and to characterize peculiar, novel
scaling at the threshold for BAM behaviour, $w=w_c$. In the interval
$0<w<w_c$ for sure LAM behaviour prevails in a whole neighbourhood of $w=0$.
However, a sort of double degeneracy of the stationary state arises
for higher $w$'s, making the scaling analysis quite difficult.
This degeneracy is an interesting phenomenon in itself, worth further
investigations, and indicates that the LAM-BAM transition
is a very complex process. 

A lower bound for $w_c$  can be estimated
by calculating the maximum amount of energy a site can receive which is needed
for at least two relaxations. Suppose all sites have the same energy $\langle \epsilon \rangle$ and 
$w_1= 1/2-w/2$ and $w_2= 1/2+w/2$. Then on adding an amount $\delta$
of energy at a site, the maximum amount of energy with which a site relaxes after the $s$-th step
of a random walk is:
$....(w_2(w_2(w_2(\delta+\langle \epsilon \rangle)+\langle \epsilon \rangle)+
\langle \epsilon \rangle))..) ...
=\langle \epsilon \rangle (1+w_2+w_2^2+w_3^3+ ..... +w_2^s)+w_2^s\delta $
which in the $s \to \infty$ limit gives $\langle \epsilon \rangle/(1-w_2)$. If this site now releases
a fraction $w_1$ of its energy, the amount left is $\langle \epsilon \rangle ((1-w_1)/(1-w_2))$
which has to be greater than one for a second relaxation to take place. Now since 
$\langle \epsilon \rangle $ can be at most
$(1-\delta)$, one must have $w_c>\delta/(2-\delta)$.

\begin{figure}[t]
\begin{center}
\includegraphics[width=6cm]{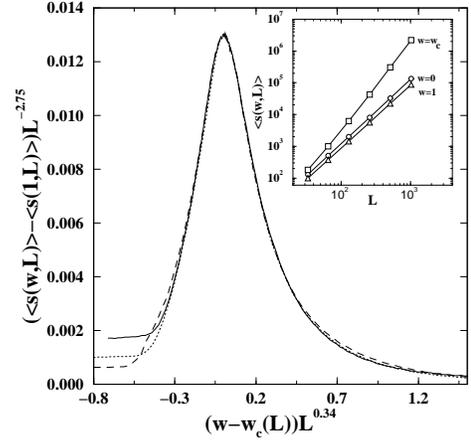}
\end{center}
\caption{
The data collapse of the difference in the average size of the avalanche $\langle s(w,L) \rangle$ and that
in the BAM $\langle s(1,L) \rangle$ is plotted for $L=65, 129$ and $257$. The inset shows
the variation of $\langle s(w,L) \rangle \sim L^{\nu_s(w)}$ where $\nu_s(w)$ = 2.75, 2.01, 2.00
for $w = w_c$, 0  and 1 respectively.
}
\label{fig4}
\end{figure}

   The basic difference in the limiting cases $w=0$ and $w=1$
is the branching of the avalanches.
Therefore, a quantity which one can monitor in order to describe the
passage from LAM to BAM behaviour is the ratio $\langle s \rangle/\langle t \rangle$.
If on average an avalanche has size $\langle s \rangle$ and life-time $\langle t \rangle$, 
$\langle s \rangle/\langle t \rangle$ 
measures the number of branches, once assumed that each branch has a 
duration $\langle t \rangle$. Therefore, we define the scaled branch number as the order parameter:
\begin{equation}
r(w,L) = (\langle s \rangle /\langle t \rangle-1)/L^{\gamma_{rs}-\gamma_{rt}}
\end{equation}
and define it as an order parameter in the limit $L \to \infty$.
By definition this order parameter must be finite and nonzero for 
$w>w_c$. It also turns out to be equal to zero for $w\le w_c$.   
Fig. 3 shows that for any $L$,
$r(w,L)$ is very close to zero for $w < w_c$, but it increases
very fast for $w > w_c$. Upon making $L$ larger, $r(w,L)$ becomes progressively
smaller for $w < w_c$, whereas it rises at a faster rate while $w > w_c$. Using 
$\delta=1/8$ we identify $w_c=0.125 \pm 0.005$. The sharp increase of $r(w,L)$
as $w \to w_c^+$ is well fitted as $r(w,L) \sim (w-w_c)^{\beta}$,
with $\beta = 1.2 \pm 0.1$.

   The average size of the avalanche $\langle s(w,L) \rangle$ has a sharp peak at
$w_c(L)$ and it depends on $L$ 
as: $w_c(L)=w_c + 0.39L^{-1/2}$ with $w_c = 0.123$. The data collapse well when 
$(\langle s(w,L) \rangle-\langle s(1,L) \rangle)L^{-2.75}$ is plotted versus $(w-w_c)L^{0.34}$. This indicates that
$\langle s(w_c,L) \rangle \sim L^{2.75}$, which implies an anomalous diffusive transport at
threshold (Fig. 4). Such anomalous transport has never been reported before for a
SOC model, to our knowledge. 
At $w=w_c$ the branch number $\langle s \rangle/\langle t  \rangle$ grows as $L^{1.31}$. So, the
threshold regime with anomalous diffusive transport is still characterized
by infinite branching of the avalanches.

   The dependence of $\langle s(w_c,L) \rangle \sim L^{2.75}$ at $w=w_c$ seems quite 
surprizing. We provide the following tentative explanation for the same:
At any nonzero value of $w$ some avalanches are purely random
walks (i.e. $s=t$) and others are branched. If $f_{RW}$ and $f_{BA}=1-f_{RW}$ are the
fractions of avalanches which are linear and branched, then for $w=0$, $f_{RW}$=1
and it decreases as $w$ increases, consequently $f_{BA}$ increases from its zero
value. Both fractions become very close to 1/2 at around $w = 0.4$ and beyond this value
they vary slowly to $f_{RW} \approx 0.6$ and $f_{BA} \approx 0.4$ at $w=1$.
We also observed the fraction of RWs which terminate within the boundary of the
system and donot drop out any energy outside the system. We find that the
fraction of such random walk avalanches increases from zero very sharply to almost 0.9 
at around $w$=0.15. This implies that around $w = w_c$ most of the 
$\delta$ input energies corresponding to those RWs which terminate within the
boundary get stored in the system. Therefore, it is the branched avalanches
which take out this extra stored energy from the system - consequently their
sizes are bigger and perhaps for this reason the average size varies as a larger
power of $L$ like 2.75.

   At $w=w_c$, the avalanche size distribution
$D(s)$ vs. $s$ has two regions with two characteristic sizes $s^1_c$ and $s^2_c$.
While $s^2_c \sim L^{2.75}$ is the usual cut-off size for the avalanches, $s^1_c \sim L^2$
is an intermediate size. For $s<s^1_c$, $\tau^1_s=0.64$ whereas
for the second region $s^1_c<s<s^2_c$ the value of $\tau^2_s$ is around 1.45. 
For $w<w_c$ but near to it, $s^1_c \sim L^2$ whereas $s^2_c$ grows as $w$ approaches $w_c$. 
However, for $w>w_c$, $s^1_c$ decreases as $s^1_c \sim (w-w_c)^{-2.25}$ as
the deviation $(w-w_c)$ increases and $\tau^2_s$ also continuously decreases to 1.3,
while $\tau^1_s \to 0$ as $L \to \infty$.

   To summarize, in a stochastic energy activation model
an active site transfers the energy to only one randomly chosen nearest neighbour.
The avalanches are linear when the transfer amount is narrow distributed and are
branched  when the transfer is broadly distributed.
While the exponents of the toppling distribution of the branchless avalanches
can be exactly determined, the universality class of the branched ones appears compatible with that of
the stochastic two-state sandpile.
A transition between the two regimes is observed by tuning the size of the window
of the stochastic transfer ratio. At the transition point, the diffusion mechanism 
induced by avalanches is anomalous.

   We thank D. Dhar and S. M. Bhattacharjee for helpful comments. S. S. M. thanks
the Dipartimento di Fisica, Universit\`a di Padova for hospitality. The work is supported
by Italian MURST-cofin99 and European Network Contract ERBFMRXCT980183. 

\leftline {E-mail: manna@boson.bose.res.in, attilio.stella@pd.infn.it}

\end{multicols}
\end {document}